\documentclass{article}
\usepackage{log_2025}						

\usepackage{booktabs}						
\usepackage{multirow}						
\usepackage{amsfonts}						
\usepackage{graphicx}						
\usepackage{duckuments}						
\usepackage{wrapfig}
\usepackage{amssymb}

\usepackage{listings}
\usepackage{xcolor}
\usepackage[most]{tcolorbox}
\usepackage{courier} 
\usepackage{lipsum}
\usepackage{siunitx}

\usepackage{caption}
\usepackage{multirow}
\usepackage{bm}
\usepackage{amsmath}
\usepackage{amsfonts}
\usepackage[colorinlistoftodos]{todonotes}
\usepackage{marginnote}

\DeclareMathOperator*{\argmax}{arg\max}

\usepackage[numbers,compress,sort]{natbib}	

\title{Tuning-Free LLM Can Build A Strong Recommender Under \\ Sparse Connectivity And Knowledge Gap Via Extracting Intent}

\author[Zheng et al.]{%
Wenqing Zheng\thanks{wenqing.zheng@capitalone.com}\\
Capital One\And
Noah Fatsi\\
Capital One\\
\And
Daniel Barcklow\\
Capital One\\
\And
Dmitri Kalaev\\
Capital One\\
\And
Steven Yao\\
Capital One\\
\And
Owen Reinert\\
Capital One\\
\And
C. Bayan Bruss\\
Capital One\\
\And
Daniele Rosa\\
Capital One
}

\begin{document}

\maketitle

\begin{abstract}

Recent advances in recommendation with large language models (LLMs) often rely on either commonsense augmentation at the item-category level or implicit intent modeling on existing knowledge graphs. However, such approaches struggle to capture grounded user intents and to handle sparsity and cold-start scenarios.
In this work, we present LLM-based Intent Knowledge Graph Recommender (IKGR), a novel framework that constructs an intent-centric knowledge graph where both users and items are explicitly linked to intent nodes extracted by a tuning-free, RAG-guided LLM pipeline. By grounding intents in external knowledge sources and user profiles, IKGR canonically represents \textit{what a user seeks} and \textit{what an item satisfies} as first-class entities. To alleviate sparsity, we further introduce a mutual-intent connectivity densification strategy, which shortens semantic paths between users and long-tail items without requiring cross-graph fusion. Finally, a lightweight GNN layer is employed on top of the intent-enhanced graph to produce recommendation signals with low latency.
Extensive experiments on public and enterprise datasets demonstrate that IKGR consistently outperforms strong baselines, particularly on cold-start and long-tail slices, while remaining efficient through a fully offline LLM pipeline.

\end{abstract}


\maketitle

\vspace{-1em}
\section{Introduction}
\vspace{-0.2em}

Modern recommender systems are expected to reason over sparse and evolving interactions while serving highly personalized needs across vast catalogs. This challenge is acute in enterprise environments, where recommendations support internal search and knowledge discovery yet must cope with heterogeneous vocabularies, domain jargon, and long-tail content \cite{verma2015improving,roy2022systematic}. Collaborative filtering and graph-based models have improved user–item representation learning \cite{koren2021advances,pei2019personalized,liu2020personalized,zuo2024inductive}, while knowledge-aware recommenders further leverage structured relations for interpretability \cite{wang2018ripplenet,wang2021knowledgefusion}. Nevertheless, their effectiveness is bounded by incomplete coverage and weak connectivity, particularly for long-tail and cold-start cases \cite{chen2021scalefree}.

To mitigate incomplete coverage and weak connectivity, a prominent direction uses LLMs to inject commonsense relations, typically \textit{complement} or \textit{substitute} links then fuses them with an existing item KG (CSRec) \cite{yang2024csrec}. This line is attractive because (i) it regularizes sparse graphs with priors that are broadly valid (jackets complement sweaters; lenses complement cameras), (ii) it is offline-friendly (relations are generated once and reused), and (iii) it improves item-side coverage where merchant metadata is incomplete. However, three limitations recur in practice.
(1) \textbf{Granularity \& intent mismatch}. Category-level commonsense smooths the space but only loosely correlates with user-specific intent. Two users who click the same \textit{camera} page may have very different intents (e.g., \textit{low-light astrophotography} vs. \textit{lightweight travel kit}). Category edges cannot capture these fine distinctions, which limits cold-start personalization.
(2) \textbf{Cross-graph fusion \& alignment risk}. CSRec-style pipelines typically align an LLM-augmented commonsense graph with an existing metadata/interaction graph. This requires ontology matching, entity resolution, and confidence thresholds. Errors here introduce structural noise that is hard to debug, and different domains require repeated re-alignment \cite{wang2021knowledgefusion}.
(3) \textbf{Temporal \& domain drift}. Off-the-shelf commonsense tends to be domain-neutral and slow-moving. Enterprise or fast-evolving domains (e.g., internal tools, niche APIs) quickly deviate; commonsense edges may become stale or irrelevant without domain grounding \cite{openreview2023llmkg,zhang2024kgllm}.

Another line remains within a pre-existing KG and learns latent mixtures over relations to explain interactions \cite{wang2021kgin,du2022hakg,chen2022modeling}. 
This method has several strengths: it's entirely symbolic and structural, avoiding the need for LLMs; it uses relation paths to capture higher-order semantics; and it doesn't require cross-graph fusion because it operates within a single KG. However, these pre-existing KG methods have several notable gaps. First, the intent \textbf{remains latent and difficult to audit}, as it's represented as a vector or a mixture over relations rather than an explicit, human-readable node. This limits explainability and hinders downstream applications like intent analytics. Additionally, the model is \textbf{bounded by the existing KG}, meaning that if the base KG is sparse or doesn't align with how users actually express themselves, the model struggles to capture missing semantics and often overfits to popular hubs, which hurts its performance on the long tail \cite{chen2021scalefree}. Finally, there is a \textbf{weak tie to external knowledge}, which means issues like polysemy and synonymy in relation labels (e.g., guide, how-to, playbook) and unstandardized domain jargon persist, as the model lacks explicit text grounding or retrieval capabilities \cite{zhao2022external}.

A third family places the LLM directly in the loop — either to synthesize interactions (augment clicks/purchases) or to produce rankings end-to-end \cite{li2023recformer,lin2023alignrec,dey2025gravity,wei2024llmrec}. This approach offers benefits such as broader semantic coverage and simplified modeling with fewer bespoke modules. However, the trade-offs are substantial. Inference-time LLM calls introduce significant \textbf{latency and cost}, complicating service-level agreements and A/B testing at scale, especially for multi-stage recommenders \cite{zhang2024kgllm}. Furthermore, synthetic interactions often lead to \textbf{distribution shift and popularity bias}, as they over-represent head items and reflect the LLM's generic priors rather than the platform's actual demand patterns; even with loss calibration, realistic long-tail fidelity is difficult to guarantee \cite{wei2024llmrec}. Finally, \textbf{governance and reproducibility} are challenging due to frequent model or LoRA updates that change behavior and varying privacy filtering across deployments, making it difficult to maintain offline-online consistency and respond to incidents in enterprise environments \cite{openreview2023llmkg}.

\begin{wrapfigure}{l}{0.48\textwidth}
\centering
\vspace{-1em}
\includegraphics[width=\linewidth, trim=0 50 0 5]{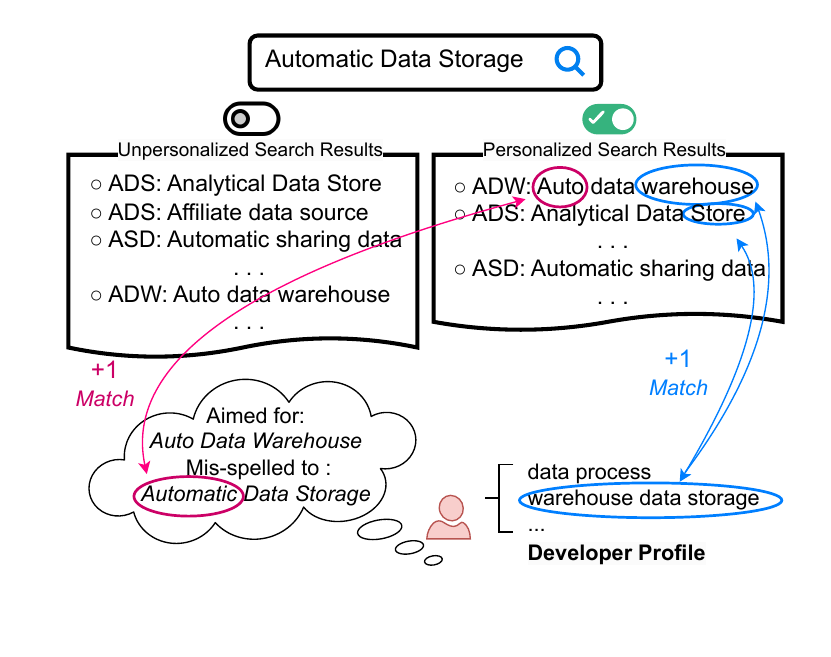}
\caption{In enterprise search, the queries could contain special terminologies and acronyms, where traditional search engines or personalized rerankers fail to capture the real intents under such knowledge gap. IKGR addresses the challenge via injecting fine-grained understanding to textual features.}
\label{fig:scenario}
\vspace{-1em}
\end{wrapfigure}
Summarizing the pros and cons of various existing methods, we distill three core challenges to LLM and KG based recommenders.

(i) User intent extraction is an effective way to clarify and densify the KG, but this step needs to be more grounded.
User profiles, queries, and enterprise documents are messy: synonyms, abbreviations, internal codenames, polysemy, and multi-lingual snippets all coexist. Extracting \textit{what a user seeks} and \textit{what an item satisfies} requires more than NER; it needs disambiguation (MLP = multilayer perceptron vs. marketing launch plan), normalization to a stable vocabulary, and aggregation across sources. Pure LLM prompts help but are prone to hallucination and label drift across batches. Without external knowledge grounding (retrieval of glossaries, wikis, policy pages) and schema-aware canonicalization, intent nodes will be noisy, redundant, and unmaintainable \cite{zhao2022external,openreview2023llmkg}.
This suggests the approach to elevate intent to first-class nodes extracted and normalized via RAG-guided LLMs.

(ii) The extracted intent needs to be more directly aligned with the existing user and items.
The goal is to shorten semantic paths by adding edges that reflect shared or similar intents (user $\leftrightarrow$ intent, item $\leftrightarrow$ intent), thereby enabling information to flow even when user–item links are missing. This step is better achieved by leveraging the extracted intents to densify a single graph, rather than through a separate cross-graph alignment stage.
If densification requires merging a separate LLM commonsense graph with an interaction/metadata graph, we re-introduce entity resolution and ontology mismatch—exactly the failure modes that cause silent errors in production \cite{yang2024csrec,wang2021knowledgefusion}.

(iii) The system should be efficient, stable, and scalable for deployment.
An efficient system should keep all heavy LLM work offline. That means handling batch extraction, incremental refresh, and backfill for new content/users; caching retrieval results; and defining compatible online components (e.g., a small GNN layer) \cite{zhang2024kgllm}. To avoid synthetic-data drift \cite{wei2024llmrec}, improvements should also come from structural connectivity and explicit intents, not from generating pseudo-interactions.

In response to these needs, we propose LLM-based Intent Knowledge Graph Recommender (IKGR), an intent-centric knowledge graph framework. First, we instantiate intent nodes and link users and items to them via a tuning-free, RAG-guided LLM extraction pipeline, grounded in external knowledge \cite{zhao2022external,openreview2023llmkg}. Second, we introduce mutual-intent connectivity densification, which structurally improves graph connectivity for sparse and cold-start regimes without relying on cross-graph fusion \cite{yang2024csrec,lin2023alignrec}. Finally, a lightweight GNN layer learns over the intent-enhanced graph to produce low-latency recommendations, fully decoupled from online LLM inference \cite{li2023recformer,zhang2024kgllm}. Empirically, IKGR consistently outperforms strong baselines on public and enterprise datasets, especially on long-tail slices, while providing interpretable pathways via explicit intents.
The contributions of IKGR can be summarized as follows.

\begin{itemize}
\item IKGR introduces an intent-centric KG construction approach that turns user and item intents into first-class nodes. Using a tuning-free, RAG-guided LLM extractor, it canonically links users and items to intents with high precision.
\item IKGR involves an intent connectivity densification step, which shortens semantic paths between users and long-tail items, improving sparsity and cold-start performance without cross-graph fusion.
\item IKGR's pipeline is offline and low-latency online by construction, and consistently outperforms state-of-the-art KG or LLM baselines on public and enterprise datasets, with notable improvements on long-tail and cold-start slices.
\end{itemize}

\vspace{-1em}
\section{Related Works}
\vspace{-0.2em}

\textbf{LLM and KG based Recommenders}.
Recent responses to sparsity and cold start fall into four camps. (1) LLM-assisted item-side augmentation injects category/attribute-level commonsense (e.g., complement/substitute) and fuses it into an item KG (CSRec) \cite{yang2024csrec}. This regularizes sparse graphs with broadly valid priors and is offline-friendly, yet it struggles to align with user-specific intent (two \textit{camera} clicks can imply very different needs), requires brittle cross-graph ontology/entity alignment, and drifts in fast-moving or enterprise domains without domain grounding \cite{wang2021knowledgefusion,openreview2023llmkg,zhang2024kgllm}. (2) KG-based implicit-intent modeling (KGIN) stays inside a single KG and explains interactions via latent mixtures over relations \cite{wang2021kgin}. It avoids fusion and can exploit multi-hop paths, but the \textit{intent} remains a non-auditable latent vector, performance is bounded by the coverage of the base KG (hurting the long tail), and weak ties to text/external knowledge leave synonymy/polysemy unresolved \cite{chen2021scalefree,zhao2022external}. (3) LLM-as-Recommender / synthetic interactions broaden semantic coverage or even replace ranking end-to-end \cite{li2023recformer,lin2023alignrec,wei2024llmrec}, but introduce inference-time cost and tail latency, amplify popularity bias in generated data, and complicate governance and reproducibility under frequent model updates \cite{zhang2024kgllm,openreview2023llmkg}. (4) Text/multimodal node enrichment improves representations by attaching reviews/UGC/features to nodes, yet leaves knowledge ungraphized—no new edges/nodes for message passing or structural densification \cite{ko2022survey,liang2023learn}. Taken together, current trends either inject coarse item-side priors, keep intent latent and hard to use beyond ranking, or pay online-LLM costs and stability penalties.

\textbf{Recommender Approaches Discovering Intent}.
Recent works explore intent modeling by leveraging external signals such as search queries and knowledge graphs. UDITSR~\cite{zhang2024unified} jointly models search and recommendation by utilizing explicit search queries to infer implicit demand intents, while a dual-intent translation mechanism captures relationships between inherent intent, demand intent, and item interactions. Alternatively, knowledge graph-based methods, such as KGIN~\cite{wang2021learning}, refine intent discovery by representing user-item interactions through fine-grained relational paths, improving interpretability and recommendation quality. These approaches demonstrate that integrating external intent signals and structured relational modeling enhances intent-aware recommendations.

\textbf{LLM-Based Interaction Augmentation for Recommender Systems}.
Gen-RecSys provides a comprehensive overview of generative models in recommendation, highlighting LLM-driven natural language understanding and multimodal integration \cite{deldjoo2024review}. LLMRec introduces graph augmentation strategies using LLMs to enrich interaction graphs, refine side information, and denoise implicit feedback, demonstrating performance gains across benchmark datasets \cite{wei2024llmrec}. BLAIR further bridges language and recommendation by pretraining sentence embeddings on large-scale review data, improving item retrieval in complex natural language contexts \cite{hou2024bridging}. Additionally, BERT4Rec employs bidirectional self-attention to model user behavior sequences, overcoming limitations of traditional sequential models \cite{sun2019bert4rec}. These studies collectively showcase the potential of LLMs in augmenting interactions, refining representations, and enhancing recommendation performance.

More related works regarding data sparsity, LLM as recommender, and personalized re-rankers are detailed in Appendix \ref{appendix:related works}.

\vspace{-1em}
\section{Preliminaries}
\label{sec:prelim}
\vspace{-0.2em}

We leverage this section to present the problem formulation and commonly adopted techniques across these domains, and introduce our contributions in Section \ref{sec:method}.

\begin{figure*}[t]
\centering
  \includegraphics[width=0.75\textwidth, trim=0 20 0 10]{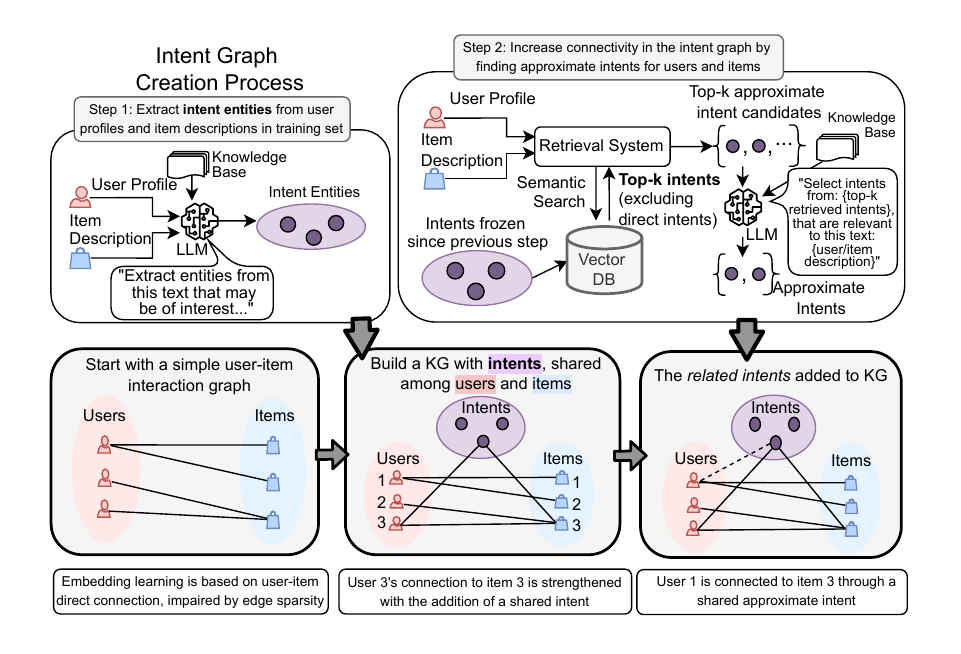}
  \caption{IKGR's graph augmentation steps to build intent knowledge graph with LLM. IKGR overcomes knowledge gap via in-context learning, avoids synthetic noise by focusing on the simple node-level intent entity retrieval task, and being light weight and tuning-free.}
  \label{fig:arch}
\vspace{-1.5em}
\end{figure*}

\vspace{-0.5em}
\subsection{Graph Formulation of Recommendation System}
We consider the recommendation system where a set of registered users $\mathcal{U}$ interact over a set of items $\mathcal{I}$. Define the interactions as a graph $\mathcal{G}$, and each user $u$ or item $i$ is a node in the graph. The collection of user-item interactions are the edges in the graph, denoted as $\mathcal{E}$. We employ the implicit feedback protocol, where each edge of $(u,i)$ implies the user $u\in \mathcal{U}$ consumes the item $i\in\mathcal{I}$. The goal is to learn a model that recommends the top-N items for a target user.

The goal is to learn a scoring function that is trained over some graph $\mathcal{G'}$, and predicts positive interactions among a set of $(u,i)$ pairs at inference time. This scoring function is denoted as \\$g_\Theta(\{(u, i)\text{ pairs}\}; \mathcal{G'})$, parameterized by $\Theta$.
In mathematical formulation, the objective of the recommendation systems is to maximize the link prediction posterior probability of accurately predicting all interactions in the dev set:

\begin{align}
    \Theta^* = \argmax\limits_{ \Theta } p( g_\Theta(\{(u, i)\}_\text{dev}; \mathcal{G}_\text{train}) = \mathcal{E}_\text{dev} ) 
\end{align}
where $\{(u, i)\}_\text{dev}$  is the dev set input user-item pairs, $\mathcal{G}_\text{train}$ is the graph used to train the model, $\mathcal{E}_\text{dev}$ is the ground truth positive edges in the dev set inputs. The formulation above means that the model maximizes the probability of accurately predicting edges on the dev set, while observing certain training graph $\mathcal{G}_\text{train}$.

We study a simple and effective approach to augment the graph, which adds new nodes and edges to augment the original graph $\mathcal{G}$ into a larger graph $\mathcal{G}\cup\mathcal{G^+}$. The training objective then becomes:

\begin{align}
    \Theta^* = \argmax\limits_{ \Theta } p( g_\Theta(\{(u, i)\}_\text{dev}; \mathcal{G}_\text{train}\cup \mathcal{G}^+) = \mathcal{E}_\text{dev} ) 
\end{align}
The augmented graph $\mathcal{G}^+$ is comprised of a unified set of intent nodes and heterogeneous connections to existing graph nodes, which is visualized in Figure \ref{fig:arch} and to be described in Section \ref{sec:method:aug}.

\vspace{-0.5em}
\subsection{Knowledge Graph Convolution Layer}

A Knowledge Graph (KG) is a directed graph composed of \textit{subject}-\textit{property}-\textit{object} triple facts. Each triplet $(e_h, e_t, r)$ denotes a relationship $r$ from head entity $e_h$ to tail entity $e_t$.

Similar to KGCN \cite{wang2019knowledge}, we employ a knowledge graph convolution layer to capture structural proximity among entities in a knowledge graph. The model learns node embeddings $\bm{E}\in R^{ N \times d}$ and relation embeddings $\bm{R}\in R^{T_R \times d}$, where $T_R$ is number of relation types, $N$ is number of nodes, and $d$ is the embedding dimensionality. Denote the input features for some node $v$ as $\mathbf{v}$, the set of entities directly connected to $v$ as $\mathcal{S}_v$, then the output embedding is the summation-aggregated neighborhood entity and relation embeddings:
\begin{equation}
\label{eq:kgcn}
    \mathbf{v}^\text{out} = \sigma \left( {\bf W} \cdot [{\bf v} + \text{softmax}(\bm{R}[\mathcal{S}_v]\bm{E}[\mathcal{S}_v]^T)\bm{E}[\mathcal{S}_v] ] + {\bf b} \right)
\end{equation}
where $\bm{R}[\mathcal{S}_v]$ and $\bm{E}[\mathcal{S}_v]$ are the relationship embedding vector set and entity embedding vector set, each containing $|\mathcal{S}_v|$ of $d$-dimensional vectors. $\bf W$ and $\bf b$ are transformation weights and a bias term, respectively, and $\sigma$ is the activation function.


\vspace{-1em}
\section{Methodology}
\label{sec:method}
The key components of IKGR involve grounded user/item intent extraction with RAG, KG densification and a GNN prediction layer.

\vspace{-0.5em}
\subsection{Intents Extraction with RAG}
\label{sec:method:aug}
\vspace{-0.5em}

Despite significant progress in incorporating side information into recommendation systems, introducing low-quality side information may even undermine what little signal we can glean from sparse interactions.
To address this challenge, the proposed IKGR focuses on an effective and simple node-level graph augmentation.
We leverage an LLM to build a knowledge graph with an additional type of node: the interaction intent entities.
The intent nodes are linked to the existing graph via two types of edges: exact intent $\mathcal{E}_{E}$ and related intent $\mathcal{E}_{R}$.

In order to address challenges posed by sparse user behavior scenarios, we leverage the LLM to perform data augmentation. Unlike existing LLM recommendation methods that directly synthesize user-item interactions \cite{wei2024llmrec}, we take a more conservative approach, tasking the LLM with a simpler, more reliable role to minimize noise in the generated outputs. This approach leverages the LLM's pre-trained common sense knowledge for accurate distillation.

We note that a forward call to the LLM with appropriate instructions is able to extract an intent entity set. For example, for item node $i$, we denote the set of intent entities as $\Omega_i$, $\Omega_i = LLM($\textit{\#\#\#Item Description: \{Item\ i\ description\} Return a list of entities mentioned in the Item Description that the user may have intents to interact with}$)$.


\subsection{Factual Knowledge Access for Grounded Extraction}
The intent extraction step assumes a high quality \textit{Item Description} paragraph describing the intent. However, the \textit{Item Description} can often be poorly-formulated in the datasets, leading to a gap in properly extracting intent. (i) In the case of enterprise search, the existence of private enterprise knowledge makes it difficult for the LLM to fully understand the context, and (ii) in the open sourced datasets case, only the item names are available without appropriate description - though details of these items are often available via online search or baked into the LLM, so that an agent with web access is able to augment it easily \cite{xi2024towards}.

We bridge this gap by feeding the intent extraction module with enterprise private knowledge or open-world knowledge.
Specifically, for our adapted enterprise use-case, user profiles and item descriptions often contain abbreviations and domain-specific concepts unknown to LLMs. We curate a knowledge base of key-value pairs, where keys are the domain-specific abbreviations/concepts, and values are their explanations. We identify any such terminology present in the user profile or item description and append their corresponding explanations to the prompt under a $\#\#\# Concept\ Explanation$ tag. This enriches the context for the LLM, enabling better understanding without the need for fine-tuning.
For open source datasets, the item name is first expanded into a paragraph summary by an LLM agent before it feeds into the entity extraction module.

\vspace{-1em}
\subsection{Intent Connectivity Enrichment via RAG}
\vspace{-0.2em}
The connectivity of the extracted intents could follow a very long-tail distribution, meaning that the majority of intent nodes might only link to few user/item nodes. To mitigate these challenges, a two-round process is used to extract and densify the intents KG. In the first round, a simple prompt template is used to generate specific entities, linking users and items to intent nodes. However, this initial graph may be sparsely connected, with many intent nodes linked to only a few users/items. To address this, a second round enriches connectivity by linking additional user/item nodes to existing intent nodes, avoiding the computationally expensive alternative of grouping similar intents ($\mathcal{O}(N^2)$ complexity)

In the second round, each user and item node is connected to additional intent entities from the fixed pool of intents generated in the first round. We call these new connections "related intents". During this second round of related intent selection, the existing intents for a given user/item node are excluded from the retrieval step.

Denote the intent entities obtained after the first round extraction as $\bar{\Omega}$. Given a user profile or an item description $text$, the second round intent extraction prompt can be formulated as: $LLM($\#\#\#Knowledge Context: ... \#\#\# Options: $R(x, \mathcal{N} \backslash \Omega_x, K)$ What are the intents mentioned in $x$ that are the most relevant?$)$ , where $R(x, \bar{\Omega} \backslash \Omega_x', K)$ retrieves a group of $K$ intents that are semantically similar to the input text $x$, yet have not been extracted during the first round output ($\Omega_x'$). The full prompt is provided in Appendix \ref{appendix:full prompt and demo}.

New users and items undergo this two-round extraction to connect them to the existing, well-connected knowledge graph.  Intent construction is always applied to items (representing an item satisfying a user intent).  When available, user profile data is also used for intent construction, and the resulting user and item intent nodes are merged using case-insensitive exact matches.  Both rounds prompt the LLM for structured output.

\vspace{-0.5em}
\subsection{Generating Recommendation Candidates with Graph Module}
\label{sec:needing more graph options}
\vspace{-0.5em}

After the KG has been built, a graph module is used to generate the recommendation candidate list. While multiple combinations of GNN architectures and loss functions could fit, we find a simple translation layer based architecture \cite{cao2019unifying} that injects the learned intents as structural priors outperforms vanilla GNN options. We leverage this intent prior GNN as the default option in the experiment sections, and discuss the details in Appendix \ref{appendix:translation layer}. We also benchmark across three other options and show results in Section \ref{sec:exp-ablation}.

\vspace{-1em}
\section{Experiments}
\vspace{-0.5em}

In this section, we verify the performance of the proposed recommender using both proprietary and open-source data \footnote{code release: \url{https://github.com/CapitalOne-Research/IKGR}}.
For the proprietary data, we introduce IKGR to an enterprise knowledge search platform to test IKGR's effectiveness in re-ranking search results. 
We also benchmark against existing baselines on four open-source recommendation datasets to verify the results in diverse scenarios. The details of these datasets are presented in Table \ref{tab:dataset}. 

We use Llama-3.1-8B model for LLM inference, and all-mpnet-base-v2 from sentence transformer for encoding textual features. 
We retrieve the top 100 prebuilt intent candidates in the kNN retrieval step of RAG.
We apply an 8:1:1 ratio when sampling the positive/negative edges for train/dev/test sets in the graph. 

\begin{table}[h]
\vspace{-2em}
\centering
\caption{Statistics of datasets: Density is the ratio of interactions over \#Users$\cdot$\#Items, \#IntEdges is the total number of connectivities between intent node and other graph nodes. AvgIntDeg is the average intent node degree.}
\resizebox{0.5\textwidth}{!}{
\label{tab:dataset}
\begin{tabular}{l r r r r r r}
\toprule[2pt]
Datasets & Search & Beauty & Books & Steam & Yelp2022 \\
\midrule
\#Users & \num{36033} & \num{40226} & \num{251394} & \num{281428} & \num{1987898}  \\
\#Items & \num{872678} & \num{54542} & \num{25606} & \num{13044} & \num{150347} \\
\#Inter & 3.5M & 0.35M & 3.2M & 3.5M & 6.9M \\
Density & 0.011\% & 0.02\% & 0.05\% & 0.095\% & 0.002\%\\
\#Intents & \num{495285} & \num{39305} & \num{45932} & \num{53940} & \num{65040} \\
\#IntEdges & 6.9M & 209K & 390K & 231K & 409K \\
AvgIntDeg & 13.9 & 5.3 & 8.5 & 4.3 & 6.3 \\
\bottomrule[2pt]
\end{tabular}
}
\end{table}

\vspace{-0.5em}
\subsection{Baselines and Datasets}

\begin{table*}[t]
\setlength{\tabcolsep}{0.82em}
    \centering
    \caption{Performance comparison of different methods. Bold scores are the best in each row, while underlined scores are the second best.} %
    \resizebox{0.88\textwidth}{!}{
    
        \begin{tabular}{l l c c c c c c c c c ccccccc}
        \toprule[2pt]
        Datasets & Metric & KGIN  & CSRec & HAKG & LLMRec & RippleNet & KGCN & KTUP & IKGR  \\
        \midrule

    \multirow{6}{*}{Search} & HR@1 & 0.0074 & 0.0079 & \underline{0.0080} & 0.0075 & 0.0024 & 0.0051 & 0.0078 & \textbf{0.0086} \\
    & HR@5 & 0.0162 & 0.0156 & \underline{0.0187} & 0.0158 & 0.0118 & 0.0161 & 0.0166 & \textbf{0.0202} \\
    & HR@10 & 0.0263 & 0.0258 & 0.0255 & 0.0253 & 0.0218 & 0.0232 & \underline{0.0262} & \textbf{0.0267} \\
    & NDCG@5 & 0.0142 & 0.0144 & 0.0139 & \underline{0.0148} & 0.0068 & 0.0101 & 0.0137 & \textbf{0.0151} \\
    & NDCG@10 & 0.0154 & 0.0153 & 0.0161 & \underline{0.0164} & 0.0087 & 0.0123 & 0.0142 & \textbf{0.0172} \\
    & MRR & 0.0135 & 0.0136 & 0.0128 & \underline{0.0143} & 0.0091 & 0.0100 & 0.0130 & \textbf{0.0153} \\
    \midrule
    \multirow{6}{*}{Beauty} & HR@1 & 0.1103 & 0.1401 & \textbf{0.1623} & 0.1563 & 0.0532 & 0.0984 & 0.1783 & \underline{0.1369} \\
    & HR@5 & 0.3092 & 0.2044 & 0.2984 & 0.2803 & 0.1972 & 0.3120 & \textbf{0.3388} & \underline{0.3316} \\
    & HR@10 & 0.4194 & 0.4293 & 0.4817 & 0.4204 & 0.3695 & 0.4204 & \underline{0.4610} & \textbf{0.4846} \\
    & NDCG@5 & 0.2398 & 0.2390 & 0.2184 & 0.2293 & 0.1307 & 0.1729 & \underline{0.2583} & \textbf{0.2806} \\
    & NDCG@10 & 0.2643 & 0.2433 & 0.2580 & 0.2930 & 0.1713 & 0.2345 & \underline{0.2814} & \textbf{0.2939} \\
    & MRR & 0.2294 & 0.2300 & 0.2402 & 0.2203 & 0.1382 & 0.2254 & \underline{0.2581} & \textbf{0.2641} \\
    \midrule
    \multirow{6}{*}{Books} & HR@1 & 0.1218 & \underline{0.1194} & 0.1177 & 0.1020 & 0.0480 & 0.0853 & 0.1125 & \textbf{0.1251} \\
    & HR@5 & 0.2983 & \underline{0.3093} & 0.2764 & 0.2674 & 0.1691 & 0.2465 & 0.2652 & \textbf{0.3197} \\
    & HR@10 & 0.3204 & 0.3449 & 0.3781 & \underline{0.4094} & 0.3553 & 0.3582 & 0.3614 & \textbf{0.4248} \\
    & NDCG@5 & 0.1910 & \underline{0.1980} & 0.1874 & 0.1877 & 0.1031 & 0.1579 & 0.1921 & \textbf{0.2097} \\
    & NDCG@10 & 0.2573 & 0.2673 & \textbf{0.2963} & 0.2673 & 0.1607 & 0.2016 & 0.2415 & \underline{0.2814} \\
    & MRR & 0.2130 & 0.2203 & \underline{0.2599} & 0.2563 & 0.1407 & 0.1832 & 0.2193 & \textbf{0.2672} \\
    \midrule
    \multirow{6}{*}{Steam} & HR@1 & 0.0783 & 0.0847 & 0.0960 & 0.0744 & 0.0304 & 0.0641 & \underline{0.1060} & \textbf{0.1095} \\
    & HR@5 & 0.2653 & 0.2483 & \underline{0.2665} & 0.2174 & 0.1429 & 0.1966 & 0.2445 & \textbf{0.2759} \\
    & HR@10 & 0.2901 & 0.2899 & 0.3170 & \underline{0.3237} & 0.2735 & 0.3218 & 0.3126 & \textbf{0.3574} \\
    & NDCG@5 & 0.1691 & 0.1335 & 0.1221 & \underline{0.1694} & 0.0915 & 0.1496 & 0.1614 & \textbf{0.1735} \\
    & NDCG@10 & 0.1739 & 0.1562 & 0.1771 & 0.1884 & 0.1161 & \underline{0.1905} & 0.1809 & \textbf{0.2212} \\
    & MRR & 0.2007 & 0.1965 & 0.1882 & 0.1872 & 0.1294 & 0.1528 & \underline{0.1957} & \textbf{0.2168} \\
    \midrule
    \multirow{6}{*}{Yelp2022} & HR@1 & 0.0771 & 0.0936 & \textbf{0.1077} & 0.0724 & 0.0495 & 0.0917 & 0.1005 & \underline{0.0989} \\
    & HR@5 & \underline{0.2766} & 0.2355 & 0.2687 & 0.2211 & 0.1513 & 0.2405 & 0.2395 & \textbf{0.2869} \\
    & HR@10 & 0.3108 & 0.3362 & \underline{0.3760} & 0.3228 & 0.2996 & 0.3395 & 0.3317 & \textbf{0.3966} \\
    & NDCG@5 & 0.1733 & 0.1823 & 0.1995 & 0.1931 & 0.1201 & 0.1397 & \underline{0.2067} & \textbf{0.2078} \\
    & NDCG@10 & 0.2034 & 0.2164 & \underline{0.2234} & 0.2127 & 0.1225 & 0.1861 & 0.2145 & \textbf{0.2277} \\
    & MRR & \underline{0.2093} & 0.2029 & 0.1842 & 0.1980 & 0.1356 & 0.1578 & 0.1906 & \textbf{0.2100} \\

        \bottomrule[2pt]
        \end{tabular}
    }
    \label{tab:result}
\end{table*}

\textbf{Enterprise Search}. In the enterprise search setting, a set of developers query the search engine to search for datasets published by other developers within the enterprise. 
The search engine first uses BM25 \cite{whissell2011improving} to retrieve a list of item candidates, then the IKGR is applied to rerank the search results. Evaluations are done over user's feedback on how high the item of interest could be ranked among the final reranked list.
In this setting, the items are enterprise datasets, which consist of text features such as dataset name, description, column names, ID labels, etc. The users are all registered within the enterprise and their developer profiles could be collected through separate channels to offer a hint about their dataset consumption preference. The IKGR is trained over historical user-dataset consumption interactions collected from separate channels. 
This dataset is labeled as \textit{Search} in dataset description table and result tables.

\vspace{-0.2em}
\textbf{Books, Beauty}. These datasets are obtained from Amazon review\footnote{\url{http://jmcauley.ucsd.edu/data/amazon/}} in \cite{DBLP:conf/sigir/McAuleyTSH15}, which contains a variety of categories. We utilize the Books and Beauty categories. We leverage the features of title, sales type, sales rank, categories, price, and brand.

\vspace{-0.2em}
\textbf{Steam}\footnote{\url{https://cseweb.ucsd.edu/~jmcauley/datasets.html\#steam_data}}. This is a dataset collected from Steam \cite{DBLP:conf/icdm/KangM18}, a large online video game distribution platform. We leverage the item features of app name, genres, publisher, sentiment, specs, tags.

\vspace{-0.2em}
\textbf{Yelp2022}\footnote{\url{https://www.yelp.com/dataset}}. This is a popular dataset for business recommendation. Given the large size, we use the transaction records after \textit{January 31st, 2022}. We treat the categories of businesses as attributes for items, and user compliment types as attributes for users.

We employ Hit Ratio (HR), Normalized Discounted Cumulative Gain (NDCG), and Mean Reciprocal Rank (MRR) as evaluation metrics.
We report HR and NDCG with $k={1,5,10}$. For all these metrics, the higher the value, the better the performance.

To verify the effectiveness of our method, we compare it with the following representative baselines.
\textbf{KGIN}~\cite{wang2021kgin} extracts non-readable intent in the latent embedding space.
\textbf{CSRec}~\cite{yang2024csrec} adds new nodes into KG with LLM, then post align the graph.
\textbf{HAKG}~\cite{du2022hakg} is a hierarchy KG built upon the hyperbolic space.
\textbf{ConvNCF}~\cite{he2018outer} employs conv layers to learn correlations in neural collaborative filtering.
\textbf{FPMC}~\cite{rendle2010factorizing} captures users' general taste as well as their sequential behaviors by combining MF with first-order Markov chains.
\textbf{LLMRec}~\cite{wei2024llmrec}, an LLM as interaction synthesizer approach.
\textbf{RippleNet}~\cite{wang2018ripplenet} uses a single layer of embedding translation by applying transH to user-item interaction.
\textbf{KGCN}~\cite{wang2019knowledge}, a KG based GNN that captures inter-item relatedness.
\textbf{KTUP}~\cite{cao2019unifying} uses a single layer of embedding translation to user-item interaction.

\vspace{-0.5em}
\subsection{Result Analysis}
\label{sec:exp-main}
\vspace{-0.5em}

Observing the results in Table \ref{tab:result}, different datasets have different levels of knowledge gaps, with the Search dataset possibly having the largest knowledge gap for the LLM. This could be indicated by the number of intents and average intent node degrees in Table \ref{tab:dataset}, where the Search dataset's intent count is about one magnitude larger than other datasets. 
From the approaches perspective, IKGR shows new state-of-the-art performances especially on the Search dataset.


\begin{wraptable}{r}{0.4\textwidth}
\vspace{-1em}
    \centering
\caption{Statistical significance test with baselines.}
\resizebox{0.4\columnwidth}{!}{
\begin{tabular}{lccc}
\toprule[2pt]
\textbf{Dataset} & \textbf{Best Baseline} & \textbf{$p$-value} & \textbf{95\% CI (IKGR)} \\
\midrule
Search & LLMRec & 0.021 & [0.0148, 0.0158] \\
Beauty & KTUP & 0.038 & [0.2605, 0.2677] \\
Books  & HAKG & 0.034 & [0.2614, 0.2730] \\
Steam  & KTUP & 0.018 & [0.2089, 0.2347] \\
Yelp2022 & KGIN & 0.061 & [0.2055, 0.2144] \\
\bottomrule[2pt]
\end{tabular}
}
\label{tab:significance}
\end{wraptable}
Each reported score in Table~\ref{tab:result} corresponds to the average of five independent runs with different random seeds. 
To assess statistically significancy, we employed a two-tailed paired $t$-test with $\alpha = 0.05$ for the per-run results of IKGR against the strongest baseline shown in Table. \ref{tab:significance}. 
The best baseline is selected per dataset according to the average MRR score.
A result is considered statistically significant if the null hypothesis (\emph{no difference between means}) is rejected at the $p<0.05$ level. 
In addition, we report $95\%$ confidence intervals computed via bootstrapping (with 10{,}000 resamples) to quantify the uncertainty of performance estimates. As shown in Table~\ref{tab:significance}, the performance gains of IKGR are statistically significant ($p < 0.05$) on four out of five datasets. Even on the \textit{Yelp2022} dataset, where margins are smaller, IKGR maintains a consistent advantage of performance gap from baselines.

\vspace{-0.5em}
\subsection{Ablation Studies}
\vspace{-0.5em}
\label{sec:exp-ablation}

To investigate the influence of each component over the enterprise search data, we conduct a list of experiments that drop each component in the proposed pipeline and compare with the original pipeline on the Search and Beauty dataset. Results are presented in Table \ref{tab:ablation}. 

\begin{table}[h]
\vspace{-1.5em}
    \centering
    \caption{Ablation Study}
    \resizebox{0.7\textwidth}{!}{
    \label{tab:ablation}
    \begin{tabular}{llccccccccccc}
    \toprule[2pt]
Dataset & Versions & HR@1 & HR@5 & HR@10 & NDCG@5 & NDCG@10 & MRR   \\
    \midrule
\multirow{6}{*}{Search} & Opt1. Int. Prior GNN &  \textbf{0.0086} & \textbf{0.0202}  & \textbf{0.0267}  & \textbf{0.0151} & \textbf{0.0172} & \textbf{0.0153} \\
& Opt2. Vanilla GNN  &  0.0078 & 0.0184  & 0.048  & 0.0132 & 0.0152 & 0.0130 \\
& Opt3. Vanilla Trans.  &  0.0082 & 0.0193  & 0.0256  & 0.0140 & 0.0164 & 0.0139 \\
& Opt4. Vanilla Scoring  &  0.0084 & 0.0199  & 0.0258  & 0.0142 & 0.0161 & 0.0137 \\
& No Related Intent  &  0.0077 & 0.0185  & 0.0254  & 0.0147 & 0.0163 & 0.0134 \\
& No Intent  &  0.0073 & 0.0175  & 0.0240  & 0.0131 & 0.0150 & 0.0125 \\
    \midrule
\multirow{6}{*}{Beauty} & Opt1. Int. Prior GNN &  \textbf{0.1369} & \textbf{0.3316}  & \textbf{0.4846}  & \textbf{0.2806} & \textbf{0.2939} & \textbf{0.2641} \\
& Opt2. Vanilla GNN  &  0.1294 & 0.3201  & 0.4299  & 0.2537 & 0.2674 & 0.2502 \\
& Opt3. Vanilla Trans.  &  0.1311 & 0.3193 & 0.4272  & 0.2519 & 0.2566 & 0.2439 \\
& Opt4. Vanilla Scoring  &  0.1332 & 0.3214 & 0.4249  & 0.2555 & 0.2579 & 0.2522 \\
& No Related Intent  &  0.1266 & 0.3284  & 0.4462  & 0.2643 & 0.2741 & 0.2536 \\
& No Intent  &  0.1183 & 0.2984  & 0.4093  & 0.2463 & 0.2453 & 0.2399 \\

    \bottomrule[2pt]
    \end{tabular}
    }
\end{table}
In Table \ref{tab:ablation}, \textit{Opt1. Int. Prior GNN} means the full IKGR version without component dropping.
\textit{Opt2. Vanilla GNN} means to use a plain GNN to make predictions over the generated user-intent-item graph.
\textit{Opt3. Vanilla Trans.} means removing both GNN and intent-aware scoring function, and only use a plain graph translation layer.
\textit{Opt4. Vanilla Scoring} means removing the intent prior scoring.
These four options correspond to four candidates in modeling the intent graph, all detailed in Appendix \ref{appendix:translation layer}.
\textit{No Related Intent} means dropping the second round of \textit{related intent} retrieval using RAG, and only exact intent edges are presented in the graph, without the related intent edges. 
\textit{No Intent} drops all intent nodes and simply use the GNN of IKGR to predict over user-item graph.

As observed in the results shown in Table \ref{tab:ablation}, the relative contributions of each components of IKGR can be estimated.
The most significant observation is that the intent edges augmentation steps as a whole helps to densify the graph almost twice, as reflected by the \#IntEdges and \#Inter in Table \ref{tab:dataset}, and the MRR performance correspondingly improved 22\% (0.0125 to 0.0153 in Search dataset).
This confirms that adding intent relations to the graph improves the IKGR performance while densifying the graph with meaningful knowledge connectivities.

The proposed intent node embedding improves the performance by offering a straightforward embedding structure between user and item vectors, hence guiding the learning procedure, as shown in Table \ref{tab:ablation} that learning the relation vectors from scratch and removing the intent based scoring both harms the performance.

\vspace{-1em}
\subsection{Cold Start Metrics}
\vspace{-0.5em}

We use the Books, Steam, Yelp2022 datasets to evaluate the cold start setting. The evaluation set is chosen to be a subset of edges whose end nodes both have less than or equal to 3 interactions (node degrees). The results are presented in Table \ref{tab:cold-start}. As can be seen from the table, IKGR achieves better performance than other methods on the tail edge set, validating its effectiveness of graph augmentation in dealing with cold start.

\begin{table}[h]
\vspace{-1.5em}
    \centering
    \caption{Performance comparison over tail edges}
    \label{tab:cold-start}
    \resizebox{0.8\linewidth}{!}{
        \begin{tabular}{l | c c c | c c c | c c c}
            \toprule[2pt]
            \textbf{Dataset} & \multicolumn{3}{c}{\textbf{Books}} & \multicolumn{3}{c}{\textbf{Steam}} & \multicolumn{3}{c}{\textbf{Yelp2022}} \\
            \cline{2-10}
            \textbf{Setting} & \textbf{HR@10} & \textbf{NDCG@10} & \textbf{MRR} & \textbf{HR@10} & \textbf{NDCG@10} & \textbf{MRR} & \textbf{HR@10} & \textbf{NDCG@10} & \textbf{MRR} \\
            \hline
            IKGR & \textbf{0.4085} & \textbf{0.2791} & \textbf{0.2630} & \textbf{0.3482} & \textbf{0.2218} & \textbf{0.2005} & \textbf{0.3684} & \textbf{0.2191} & \textbf{0.1904} \\
            ConvNCF & 0.2699 & 0.1373 & 0.1382 & 0.2546 & 0.1548 & 0.1225 & 0.2436 & 0.1256 & 0.1194 \\
            FPMC & 0.2810 & 0.1427 & 0.1326 & 0.2897 & 0.2010 & 0.1453 & 0.2573 & 0.1634 & 0.1429 \\
            KTUP & 0.3114 & 0.2044 & 0.1679 & 0.3340 & 0.2186 & 0.1898 & 0.3104 & 0.1944 & 0.1774 \\
            \toprule[2pt]
        \end{tabular}
    }
\end{table}

\begin{wraptable}{r}{0.5\textwidth}
\vspace{-3em}
    \centering
    \caption{Results on hyperparameter configurations}
    \label{tab:hparam}
    \resizebox{0.5\textwidth}{!}{
        \begin{tabular}{lccc}
            \toprule[2pt]
            \textbf{Configuration} & \textbf{HR@10} & \textbf{NDCG@10} & \textbf{MRR} \\
            \hline
            $k=120$, \#conv$=1$ & 0.4262 & 0.2835 & 0.2711 \\
            $k=100$, \#conv$=1$ & 0.4248 & 0.2814 & 0.2672 \\
            $k=80$, \#conv$=1$ & 0.3610 & 0.2619 & 0.2520 \\
            $k=50$, \#conv$=1$ & 0.3023 & 0.2205 & 0.1945 \\
            $k=100$, \#conv$=2$ & 0.3735 & 0.2517 & 0.2482 \\
            \toprule[2pt]
        \end{tabular}
    }
\end{wraptable}
\vspace{-1em}

\vspace{-1em}
\subsection{Hyperparameter Sensitiveness}
\vspace{-0.5em}

To test how model behaves under different hyperparameters, we computed top-k in the kNN retrieval step of RAG, and number of GNN layers in the model architecture. These experiments are conducted on the Books dataset.

The hyperparameter sensitivity results are shown in Table \ref{tab:hparam}. As seen from the table, the model kNN saturates at k=100, which validates our architecture choice of $k$ that both ensures performance and avoid over lengthy token sequences.


\vspace{-1em}
\subsection{Impact of Knowledge Base}
\vspace{-0.5em}

\begin{wrapfigure}{r}{0.45\textwidth}
\vspace{-3em}
    \centering
    \captionof{table}{Statistics of extracted intents}
    \label{tab:kb-impact}
    \resizebox{\linewidth}{!}{
        \begin{tabular}{lS[table-format=2.3]}
            \toprule[2pt]
            \textbf{Metric} & \textbf{Value} \\
            \hline
            Average num entity with KB & 15.5 \\
            Average num entity without KB & 14.5 \\
            Avg Jaccard Similarity Coefficient & 0.892 \\
            \toprule[2pt]
        \end{tabular}
    }
\end{wrapfigure}

To quantify the influence of the knowledge base, we use the enterprise search data to compare the extracted intent sets for two scenarios: the knowledge base appended and dropped. We computed the average number of entity extracted, and the Jaccard Similarity Coefficient ( $J(A,B) = \frac{|A \cap B|}{|A \cup B|}$ ) for the entities under these two settings in Table \ref{tab:kb-impact}.


The result shows that the average number of intents does not differ much, and the Jaccard Similarity Coefficient is close to 1, meaning the knowledge base has limited impact on the extracted entity set. Our intention with the knowledge base is to serve as a lightweight information source in the event of heavy domain gap, rather than a bottleneck key component. Indeed, we have made the LLM generation task as a simple entity extraction task, hence the impact of missing structured knowledge is minimized.

\vspace{-1em}
\section{Conclusions}
\vspace{-0.5em}

In this work, we propose IKGR, a knowledge graph based recommender built with a Large Language Model (LLM). The proposed method features a data augmentation step to explicitly extract entities that the users have intents to interact with, and learns node embeddings over the knowledge graph using an embedding translation layer to combine the intent structure knowledge. 
This work takes the enterprise search personalization as a case study, and verifies that (1) when knowledge gap exists, using a simplified node-level augmentation task helps learn embeddings, while synthesizing interactions harms the model performance and introduces noise; (2) injecting intent structure prior into the modeling helps better capturing the semantic structure and boosts embedding learning.

\bibliographystyle{unsrtnat}
\bibliography{reference}

@inproceedings{xi2024towards,
  title={Towards open-world recommendation with knowledge augmentation from large language models},
  author={Xi, Yunjia and Liu, Weiwen and Lin, Jianghao and Cai, Xiaoling and Zhu, Hong and Zhu, Jieming and Chen, Bo and Tang, Ruiming and Zhang, Weinan and Yu, Yong},
  booktitle={Proceedings of the 18th ACM Conference on Recommender Systems},
  pages={12--22},
  year={2024}
}

@inproceedings{chen2022modeling,
  title={Modeling scale-free graphs with hyperbolic geometry for knowledge-aware recommendation},
  author={Chen, Yankai and Yang, Menglin and Zhang, Yingxue and Zhao, Mengchen and Meng, Ziqiao and Hao, Jianye and King, Irwin},
  booktitle={Proceedings of the fifteenth ACM international conference on web search and data mining},
  pages={94--102},
  year={2022}
}

@inproceedings{du2022hakg,
  title={HAKG: Hierarchy-aware knowledge gated network for recommendation},
  author={Du, Yuntao and Zhu, Xinjun and Chen, Lu and Zheng, Baihua and Gao, Yunjun},
  booktitle={Proceedings of the 45th international ACM SIGIR conference on Research and development in Information Retrieval},
  pages={1390--1400},
  year={2022}
}

@article{yang2024csrec,
  title   = {Common Sense Enhanced Knowledge-based Recommendation with Large Language Model},
  author  = {Shenghao Yang and Weizhi Ma and Peijie Sun and Min Zhang and Qingyao Ai and Yiqun Liu and Mingchen Cai},
  journal = {arXiv preprint arXiv:2403.18325},
  year    = {2024},
  url     = {https://arxiv.org/abs/2403.18325}
}

@inproceedings{wang2021kgin,
  author    = {Xiang Wang and Tinglin Huang and Dingxian Wang and Yancheng Yuan and Zhenguang Liu and Xiangnan He and Tat{-}Seng Chua},
  title     = {Learning Intents behind Interactions with Knowledge Graph for Recommendation},
  booktitle = {Proceedings of The Web Conference (WWW)},
  pages     = {878--887},
  publisher = {ACM / IW3C2},
  year      = {2021},
  url       = {https://arxiv.org/pdf/2102.07057}
}

@article{chen2021scalefree,
  author  = {Yankai Chen and Menglin Yang and Yingxue Zhang and Mengchen Zhao and Ziqiao Meng and Jian Hao and Irwin King},
  title   = {Modeling Scale-free Graphs for Knowledge-aware Recommendation},
  journal = {arXiv preprint arXiv:2108.06468},
  year    = {2021},
  url     = {https://arxiv.org/abs/2108.06468}
}

@article{zhao2022external,
  title   = {Learning from Rich External Knowledge for Recommendation},
  author  = {Zhao, <first name> and others},
  journal = {arXiv preprint arXiv:2204.04959},
  year    = {2022},
  url     = {https://arxiv.org/abs/2204.04959}
}

@article{wang2021knowledgefusion,
  title   = {Knowledge-aware Recommendation with Graph Neural Networks: A Survey},
  author  = {Wang, Hongwei and others},
  journal = {arXiv preprint arXiv:2112.14936},
  year    = {2021},
  url     = {https://arxiv.org/abs/2112.14936}
}

@article{lin2023alignrec,
  title   = {AlignRec: Alignment-enhanced Recommendation with Large Language Models},
  author  = {Lin, <first name> and others},
  journal = {arXiv preprint arXiv:2306.10933},
  year    = {2023},
  url     = {https://arxiv.org/abs/2306.10933}
}

@article{li2023recformer,
  title   = {RecFormer: Large Language Model for Recommendation},
  author  = {Li, <first name> and others},
  journal = {arXiv preprint arXiv:2305.07001},
  year    = {2023},
  url     = {https://arxiv.org/abs/2305.07001}
}

@inproceedings{openreview2023llmkg,
  title     = {Large Language Models for Knowledge Graph Construction: A Survey},
  author    = {Author, <first name> and others},
  booktitle = {International Conference on Learning Representations (ICLR), OpenReview},
  year      = {2023},
  url       = {https://openreview.net/pdf?id=MipDf3C38E}
}

@article{zhang2024kgllm,
  title   = {LLM-enhanced Knowledge Graphs for Recommendation: Opportunities and Challenges},
  author  = {Zhang, <first name> and others},
  journal = {arXiv preprint arXiv:2402.13840},
  year    = {2024},
  url     = {https://arxiv.org/abs/2402.13840}
}

@article{roy2022systematic,
  title={A systematic review and research perspective on recommender systems},
  author={Roy, Deepjyoti and Dutta, Mala},
  journal={Journal of Big Data},
  volume={9},
  number={1},
  pages={59},
  year={2022},
  publisher={Springer}
}

@article{verma2015improving,
  title={Improving scalability of personalized recommendation systems for enterprise knowledge workers},
  author={Verma, Chetan and Hart, Michael and Bhatkar, Sandeep and Parker-Wood, Aleatha and Dey, Sujit},
  journal={IEEE Access},
  volume={4},
  pages={204--215},
  year={2015},
  publisher={IEEE}
}

@inproceedings{wang2021learning,
  title={Learning intents behind interactions with knowledge graph for recommendation},
  author={Wang, Xiang and Huang, Tinglin and Wang, Dingxian and Yuan, Yancheng and Liu, Zhenguang and He, Xiangnan and Chua, Tat-Seng},
  booktitle={Proceedings of the web conference 2021},
  pages={878--887},
  year={2021}
}

@article{koren2021advances,
  title={Advances in collaborative filtering},
  author={Koren, Yehuda and Rendle, Steffen and Bell, Robert},
  journal={Recommender systems handbook},
  pages={91--142},
  year={2021},
  publisher={Springer}
}

@inproceedings{zhou2020s3,
  title={S3-rec: Self-supervised learning for sequential recommendation with mutual information maximization},
  author={Zhou, Kun and Wang, Hui and Zhao, Wayne Xin and Zhu, Yutao and Wang, Sirui and Zhang, Fuzheng and Wang, Zhongyuan and Wen, Ji-Rong},
  booktitle={Proceedings of the 29th ACM international conference on information \& knowledge management},
  pages={1893--1902},
  year={2020}
}

@inproceedings{sun2019bert4rec,
  title={BERT4Rec: Sequential recommendation with bidirectional encoder representations from transformer},
  author={Sun, Fei and Liu, Jun and Wu, Jian and Pei, Changhua and Lin, Xiao and Ou, Wenwu and Jiang, Peng},
  booktitle={Proceedings of the 28th ACM international conference on information and knowledge management},
  pages={1441--1450},
  year={2019}
}

@article{hou2024bridging,
  title={Bridging language and items for retrieval and recommendation},
  author={Hou, Yupeng and Li, Jiacheng and He, Zhankui and Yan, An and Chen, Xiusi and McAuley, Julian},
  journal={arXiv preprint arXiv:2403.03952},
  year={2024}
}

@article{he2018outer,
  title={Outer product-based neural collaborative filtering},
  author={He, Xiangnan and Du, Xiaoyu and Wang, Xiang and Tian, Feng and Tang, Jinhui and Chua, Tat-Seng},
  journal={arXiv preprint arXiv:1808.03912},
  year={2018}
}

@inproceedings{rendle2010factorizing,
  title={Factorizing personalized markov chains for next-basket recommendation},
  author={Rendle, Steffen and Freudenthaler, Christoph and Schmidt-Thieme, Lars},
  booktitle={Proceedings of the 19th international conference on World wide web},
  pages={811--820},
  year={2010}
}

@inproceedings{DBLP:conf/sigir/McAuleyTSH15,
  author    = {J. J. McAuley and
               C. Targett and
               Q. Shi and
               A. van den Hengel},
  title     = {Image-Based Recommendations on Styles and Substitutes},
  booktitle = {{SIGIR} 2015},
  pages     = {43--52},
  year      = {2015},
}

@inproceedings{cao2019unifying,
  title={Unifying knowledge graph learning and recommendation: Towards a better understanding of user preferences},
  author={Cao, Yixin and Wang, Xiang and He, Xiangnan and Hu, Zikun and Chua, Tat-Seng},
  booktitle={The world wide web conference},
  pages={151--161},
  year={2019}
}

@inproceedings{wang2014knowledge,
  title={Knowledge graph embedding by translating on hyperplanes},
  author={Wang, Zhen and Zhang, Jianwen and Feng, Jianlin and Chen, Zheng},
  booktitle={Proceedings of the AAAI conference on artificial intelligence},
  volume={28},
  number={1},
  year={2014}
}

@article{bordes2013translating,
  title={Translating embeddings for modeling multi-relational data},
  author={Bordes, Antoine and Usunier, Nicolas and Garcia-Duran, Alberto and Weston, Jason and Yakhnenko, Oksana},
  journal={Advances in neural information processing systems},
  volume={26},
  year={2013}
}

@inproceedings{wang2019knowledge,
  title={Knowledge graph convolutional networks for recommender systems},
  author={Wang, Hongwei and Zhao, Miao and Xie, Xing and Li, Wenjie and Guo, Minyi},
  booktitle={The world wide web conference},
  pages={3307--3313},
  year={2019}
}

@inproceedings{liang2023learn,
  title={Learn from relational correlations and periodic events for temporal knowledge graph reasoning},
  author={Liang, Ke and Meng, Lingyuan and Liu, Meng and Liu, Yue and Tu, Wenxuan and Wang, Siwei and Zhou, Sihang and Liu, Xinwang},
  booktitle={Proceedings of the 46th International ACM SIGIR Conference on Research and Development in Information Retrieval Conference on Research and Development in Information Retrieval},
  pages={1559--1568},
  year={2023}
}

@article{ko2022survey,
  title={A survey of recommendation systems: recommendation models, techniques, and application fields},
  author={Ko, Hyeyoung and Lee, Suyeon and Park, Yoonseo and Choi, Anna},
  journal={Electronics},
  volume={11},
  number={1},
  pages={141},
  year={2022},
  publisher={MDPI}
}

@inproceedings{wang2018ripplenet,
  title={Ripplenet: Propagating user preferences on the knowledge graph for recommender systems},
  author={Wang, Hongwei and Zhang, Fuzheng and Wang, Jialin and Zhao, Miao and Li, Wenjie and Xie, Xing and Guo, Minyi},
  booktitle={Proceedings of the 27th ACM international conference on information and knowledge management},
  pages={417--426},
  year={2018}
}

@inproceedings{wei2024llmrec,
  title={Llmrec: Large language models with graph augmentation for recommendation},
  author={Wei, Wei and Ren, Xubin and Tang, Jiabin and Wang, Qinyong and Su, Lixin and Cheng, Suqi and Wang, Junfeng and Yin, Dawei and Huang, Chao},
  booktitle={Proceedings of the 17th ACM International Conference on Web Search and Data Mining},
  pages={806--815},
  year={2024}
}

@inproceedings{DBLP:conf/icdm/KangM18,
  author    = {W.{-}C. Kang and
               J. J. McAuley},
  title     = {Self-Attentive Sequential Recommendation},
  booktitle = {{ICDM} 2018},
  pages     = {197--206},
  year      = {2018},
}

@article{whissell2011improving,
  title={Improving document clustering using Okapi BM25 feature weighting},
  author={Whissell, John S and Clarke, Charles LA},
  journal={Information retrieval},
  volume={14},
  pages={466--487},
  year={2011},
  publisher={Springer}
}

@inproceedings{zhang2024unified,
  title={Unified Dual-Intent Translation for Joint Modeling of Search and Recommendation},
  author={Zhang, Yuting and Wu, Yiqing and Han, Ruidong and Sun, Ying and Zhu, Yongchun and Li, Xiang and Lin, Wei and Zhuang, Fuzhen and An, Zhulin and Xu, Yongjun},
  booktitle={Proceedings of the 30th ACM SIGKDD Conference on Knowledge Discovery and Data Mining},
  pages={6291--6300},
  year={2024}
}

@inproceedings{deldjoo2024review,
  title={A review of modern recommender systems using generative models (gen-recsys)},
  author={Deldjoo, Yashar and He, Zhankui and McAuley, Julian and Korikov, Anton and Sanner, Scott and Ramisa, Arnau and Vidal, Ren{\'e} and Sathiamoorthy, Maheswaran and Kasirzadeh, Atoosa and Milano, Silvia},
  booktitle={Proceedings of the 30th ACM SIGKDD Conference on Knowledge Discovery and Data Mining},
  pages={6448--6458},
  year={2024}
}

@inproceedings{liang2010connecting,
  title={Connecting users and items with weighted tags for personalized item recommendations},
  author={Liang, Huizhi and Xu, Yue and Li, Yuefeng and Nayak, Richi and Tao, Xiaohui},
  booktitle={Proceedings of the 21st ACM conference on Hypertext and hypermedia},
  pages={51--60},
  year={2010}
}

@inproceedings{liu2020personalized,
  title={Personalized Re-ranking with Item Relationships for E-commerce},
  author={Liu, Weiwen and Liu, Qing and Tang, Ruiming and Chen, Junyang and He, Xiuqiang and Heng, Pheng Ann},
  booktitle={Proceedings of the 29th ACM International Conference on Information \& Knowledge Management},
  pages={925--934},
  year={2020}
}

@inproceedings{pei2019personalized,
  title={Personalized re-ranking for recommendation},
  author={Pei, Changhua and Zhang, Yi and Zhang, Yongfeng and Sun, Fei and Lin, Xiao and Sun, Hanxiao and Wu, Jian and Jiang, Peng and Ge, Junfeng and Ou, Wenwu and others},
  booktitle={Proceedings of the 13th ACM conference on recommender systems},
  pages={3--11},
  year={2019}
}

@inproceedings{wu2024coral,
  title={Coral: collaborative retrieval-augmented large language models improve long-tail recommendation},
  author={Wu, Junda and Chang, Cheng-Chun and Yu, Tong and He, Zhankui and Wang, Jianing and Hou, Yupeng and McAuley, Julian},
  booktitle={Proceedings of the 30th ACM SIGKDD Conference on Knowledge Discovery and Data Mining},
  pages={3391--3401},
  year={2024}
}

@inproceedings{bao2023tallrec,
  title={Tallrec: An effective and efficient tuning framework to align large language model with recommendation},
  author={Bao, Keqin and Zhang, Jizhi and Zhang, Yang and Wang, Wenjie and Feng, Fuli and He, Xiangnan},
  booktitle={Proceedings of the 17th ACM Conference on Recommender Systems},
  pages={1007--1014},
  year={2023}
}

@inproceedings{kim2024large,
  title={Large language models meet collaborative filtering: An efficient all-round llm-based recommender system},
  author={Kim, Sein and Kang, Hongseok and Choi, Seungyoon and Kim, Donghyun and Yang, Minchul and Park, Chanyoung},
  booktitle={Proceedings of the 30th ACM SIGKDD Conference on Knowledge Discovery and Data Mining},
  pages={1395--1406},
  year={2024}
}

@inproceedings{evnine2024achieving,
  title={Achieving a Better Tradeoff in Multi-stage Recommender Systems through Personalization},
  author={Evnine, Ariel and Ioannidis, Stratis and Kalimeris, Dimitris and Kalyanaraman, Shankar and Li, Weiwei and Nir, Israel and Sun, Wei and Weinsberg, Udi},
  booktitle={Proceedings of the 30th ACM SIGKDD Conference on Knowledge Discovery and Data Mining},
  pages={4939--4950},
  year={2024}
}

@inproceedings{zuo2024inductive,
  title={Inductive Modeling for Realtime Cold Start Recommendations},
  author={Zuo, Chandler and Castaldo, Jonathan and Zhu, Hanqing and Zhang, Haoyu and Liu, Ji and Ou, Yangpeng and Kong, Xiao},
  booktitle={Proceedings of the 30th ACM SIGKDD Conference on Knowledge Discovery and Data Mining},
  pages={6400--6409},
  year={2024}
}

@article{zheng2021cold,
  title={Cold brew: Distilling graph node representations with incomplete or missing neighborhoods},
  author={Zheng, Wenqing and Huang, Edward W and Rao, Nikhil and Katariya, Sumeet and Wang, Zhangyang and Subbian, Karthik},
  journal={arXiv preprint arXiv:2111.04840},
  year={2021}
}

@article{dey2025gravity,
  title={GRAVITY: A Framework for Personalized Text Generation via Profile-Grounded Synthetic Preferences},
  author={Dey, Priyanka and Rosa, Daniele and Zheng, Wenqing and Barcklow, Daniel and Zhao, Jieyu and Ferrara, Emilio},
  journal={arXiv preprint arXiv:2510.11952},
  year={2025}
}


\newpage
\appendix

\section{Full Prompts And Visualizations}
\label{appendix:full prompt and demo}

For the enterprise search scenario, the full prompt in the first round of intent generation is provided as follows.

\begin{lstlisting}[
  frame=single,
  backgroundcolor=\color{gray!10},
  breaklines=true, % 启用自动换行
  breakatwhitespace=true, % 仅在空白处换行
  columns=fullflexible, % 让字符宽度可调整，以适应对齐
  basicstyle=\ttfamily,
  xleftmargin=10pt, % 可选：调整左边距
  xrightmargin=10pt, % 可选：调整右边距
  tabsize=2, % 可选：设置 tab 键大小
]
### Task
You are given a short paragraph of description of a dataset. Extract a few entities mentioned in the description that the users may be interested in when they consume this dataset. Acronyms can also be included in your answer, if any. Acronyms connected via underscores can be broken apart and partially used, so as to only include the core part into your answer. Please discard entities that have too general and broad meaning, and only pick those specific to this dataset. Please also discard entities that are over-specific, such as date or seemingly highly specified serial numbers/markers.
Please return answer in the format of python list of string, such as: ["answer 1", "answer 2"]. Only return the python list of string. Do NOT return any other explaination words.
### Dataset Description:
{desc}
### Answer:
\end{lstlisting}

The full prompt in the second round of intent linkage densification is provided as follows.

\begin{lstlisting}[
  frame=single,
  backgroundcolor=\color{gray!10},
  breaklines=true, % 启用自动换行
  breakatwhitespace=true, % 仅在空白处换行
  columns=fullflexible, % 让字符宽度可调整，以适应对齐
  basicstyle=\ttfamily,
  xleftmargin=10pt, % 可选：调整左边距
  xrightmargin=10pt, % 可选：调整右边距
  tabsize=2, % 可选：设置 tab 键大小
]
You are given a short paragraph of description of a dataset, and a list of entity candidates. Based on the available entity candidates, which of them are likely to attract the user's attention, if the user has proven interest in the dataset? Please include related entities, which may not be diretly mentioned in the dataset description. Please make selections only from the available options given below. Please return answer in the format of python list of string.
### Dataset Description:
{desc}
### Available Options:
{RAG_options}
### Important Notes:
1. You can ONLY choose from the Available Options above. If certain entity is not mentioned in the Available Options, you CANNOT return that entity.
2. Please return as many related entities as possible.
3. Please only return the python list of string. Do NOT return any other explaination words.
### Answer:
\end{lstlisting}

For the open source dataset (e.g. Steam), the prompt for description summarization using the open world knowledge agent is as follows. 

\begin{lstlisting}[
  frame=single,
  backgroundcolor=\color{gray!10},
  breaklines=true, % 启用自动换行
  breakatwhitespace=true, % 仅在空白处换行
  columns=fullflexible, % 让字符宽度可调整，以适应对齐
  basicstyle=\ttfamily,
  xleftmargin=10pt, % 可选：调整左边距
  xrightmargin=10pt, % 可选：调整右边距
  tabsize=2, % 可选：设置 tab 键大小
]
{item_name} is a game on Steam. Collect necessary information for this game, and write one paragraph summary of why a user is potentially interested in it. Use objective tone.
\end{lstlisting}

The above prompt converts the item name into a paragraph of LLM summary. Then LLM is used to extract the intent using a similar prompt as above. An example of extracted intents for steam dataset is as follows.

\newpage

\begin{lstlisting}[
  frame=single,
  backgroundcolor=\color{gray!10},
  breaklines=true, % 启用自动换行
  breakatwhitespace=true, % 仅在空白处换行
  columns=fullflexible, % 让字符宽度可调整，以适应对齐
  basicstyle=\ttfamily,
  xleftmargin=10pt, % 可选：调整左边距
  xrightmargin=10pt, % 可选：调整右边距
  tabsize=2, % 可选：设置 tab 键大小
]
Origin item name: Half-Life 2
LLM summary: Half-Life 2 is a landmark first-person shooter that remains highly relevant today for its exceptional storytelling, physics-based gameplay, and rich world-building. Players could be interested in it because it offers more than just combat; the game's silent protagonist, Gordon Freeman, navigates a deeply immersive narrative that unfolds entirely through gameplay, pulling you into a desperate human resistance against an alien empire. The revolutionary Gravity Gun allows for creative problem-solving and combat by manipulating objects, while the game's detailed environments and memorable characters-like the resourceful Alyx Vance-create a compelling and lasting experience that set a new standard for the genre. Ultimately, it's a masterclass in game design that is both historically significant and still incredibly fun to play.
Extracted Intent entities: ["storytelling", "physics-based gameplay", "world-building", "combat", "narrative", "human resistance", "alien empire", "gravity gun", "problem-solving", "characters", "alyx vance", "game design"]
\end{lstlisting}

\section{Intent Prior GNN}
\label{appendix:translation layer}

Below we specify the details of intent prior GNN and how it improves scoring with intent awareness at test time.

\subsection{Knowledge Graph Embedding Translation Formulations}
Due to the incomplete nature of KGs, KG completion is often leveraged as a self-supervised learning task, which predicts the missing entity $e_h$ or $e_t$ for a triplet $(e_h, e_t, r)$.
To this end, TransE~\cite{bordes2013translating}, a popular knowledge graph embedding model commonly used for KG completion, enforces a translation in the embedding space: $\mathbf{e}_h+\mathbf{e}_r\approx \mathbf{e}_t$.
This embedding translation is achieved via the following training objective: $\min\mathop{\Sigma}\limits_{\mathbf{e}_h, \mathbf{r}, \mathbf{e}_t } \parallel \mathbf{e}_h+\mathbf{r}-\mathbf{e}_t \parallel$.

One issue with TransE embedding is that a single relation type may correspond to multiple head entities or tail entities, leading to significant 1-to-N, N-to-1, and N-to-N issues~\cite{wang2014knowledge}. As an improvement, TransH~\cite{wang2014knowledge} learns different representations for an entity conditioned on different relations. It assumes that each relation owns a hyperplane, and the translation between head entity and tail entity is valid only if they are projected to the same hyperplane. It defines an energy score function for a triplet as follows:

\begin{equation}
f(\mathbf{e}_h, \mathbf{e}_t, \mathbf{r})=\parallel \mathbf{e}_h^{\bot}+\mathbf{r}-\mathbf{e}_t^{\bot} \parallel
\end{equation}
where a lower score of $f(e_h, e_t, r)$ indicates that the triplet is likely valid. $\mathbf{e}_h^{\bot}$ and $\mathbf{e}_t^{\bot}$ are projected entity vectors:

\begin{equation}
\mathbf{e}_h^{\bot}=\mathbf{e}_h-\mathbf{w}_r^{\mathrm{T}} \mathbf{e}_h \mathbf{w}_r
\end{equation}

\begin{equation}
  \mathbf{e}_t^{\bot}=\mathbf{e}_t-\mathbf{w}_r^{\mathrm{T}} \mathbf{e}_t \mathbf{w}_r
\end{equation}

\noindent where $\mathbf{w}_r$ and $\mathbf{r}$ are two learned vectors for a specified relation $r$. $\mathbf{w}_r$ denotes the projection vector of its corresponding hyperplane where  $\mathbf{r}$ is the relationship embedding vector.

Finally, the training of TransH encourages the discrimination between valid triplets and incorrect ones using margin-based ranking loss. Similar to KTUP~\cite{cao2019unifying}, we employ:

\begin{equation}
\label{eq:loss:kg}
  \mathcal{L}_{KG}=-\sum_{\substack{(\mathbf{e}_h, \mathbf{e}^+_t, \mathbf{r}^+)\in\mathcal{KG} \\ (\mathbf{e}_h, \mathbf{e}^-_t, \mathbf{r}^-)\in\mathcal{KG}^-}} \log\sigma[f(\mathbf{e}_h, \mathbf{e}^-_t, \mathbf{r}^-)-f(\mathbf{e}_h, \mathbf{e}^+_t, \mathbf{r}^+)]
\end{equation}

\noindent where $\mathcal{KG}^-$ contains corrupted triplets constructed by randomly sampling a tail entity and relation. In practice, weight decay and normalization-enforcing losses are also applied to prevent overfitting.

\begin{figure*}
  \includegraphics[width=0.95\textwidth, trim=0 20 0 20]{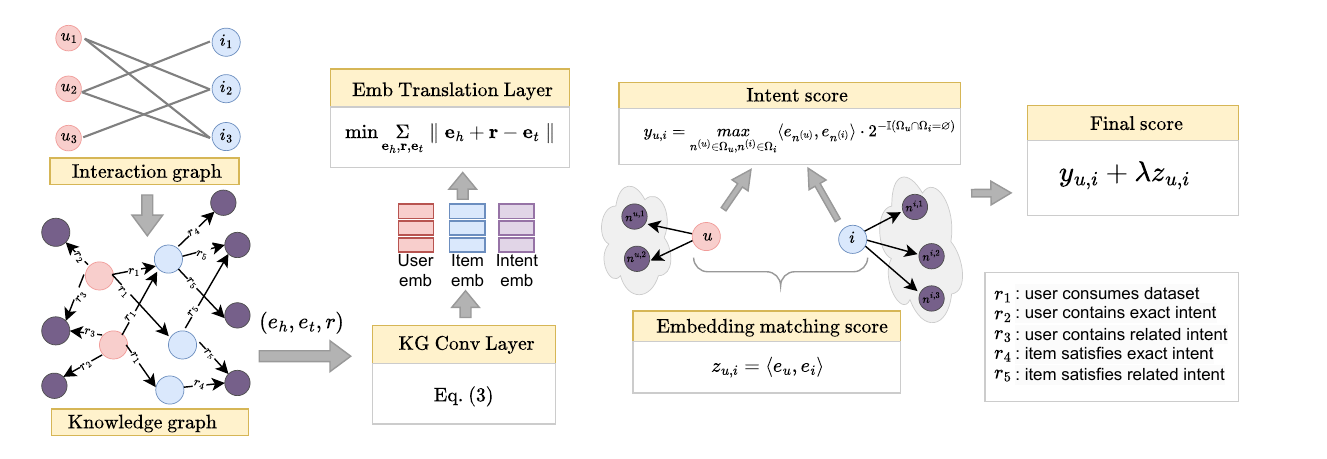}
  \caption{The intent prior GNN in IKGR. This module is the knowledge graph convolution layer followed by an embedding translation layer, where the intent translation is encoded as structure prior in the embedding space.}
  \label{fig:kgnn}
\end{figure*}

\subsection{Intent Prior Injected Translation layers}
Having obtained the intent nodes from items and users, we train a knowledge graph-based GNN to predict user-item interaction probabilities. The architecture is shown in Figure \ref{fig:kgnn}, which is a knowledge graph convolution layer followed by a transH layer.
We note that this component of the system can be swapped for any KG-compatible graph learning method. 
We posit that as long as the node-level features are efficiently decoupled and connections are made, semantic meaning can be effectively captured even with a relatively simple graph model.
Therefore, we leverage a simple embedding translation-based approach to verify that building intent-item and intent-user connections helps enable more accurate user-item interaction predictions even with simpler architectures.
This GNN module is implemented for the ease of the benchmarking and ablation study, so that the gain from the intent graph augmentation can be readily observed.
Next, we discuss the embedding translation mechanisms and how we connect the user-item preference translation with the newly added intent embeddings at inference time.

First, we leverage a pre-trained natural language encoder to process textual node features into embeddings as input to IKGR. Given the raw embedding input, we apply a KG convolution layer described in Equation \ref{eq:kgcn} to pass signals among intent, item, and user nodes to prepare for interaction prediction. Then, the KG embedding translation layer is applied to encourage semantic meaning alignment given the identified intent nodes in the graph.

In the intent-augmented knowledge graph, there are three main types of relations: user possesses intent, item satisfies intent, and user consumes item. For the first two relation types, we leverage Equation \ref{eq:loss:kg} to train relation representation $\mathbf{r}$ and projection vector $\mathbf{w}_r$ without modification. For the third type, user consumes item, the projection vector $\mathbf{w}_r$ is still independently trained, but the translation vector $\mathbf{r}$ is based on intent node embeddings. 

The motivation behind leveraging the intent nodes to compute relation representation is that both the user and the item are decoupled into lists of intents, and when there are shared or similar intent nodes, their difference is expected to be small, i.e., $\mathbf{e}_h^{\bot} \approx \mathbf{e}_t^{\bot}$ or $\parallel \mathbf{r}\parallel\approx0$. In this way, the intent embedding introduces direct insight for the relation embedding translations. We hence build the relation embeddings $\mathbf{r}^{u,i}\in R^d$ as follows.

Denote the intent embeddings for user node $u$ and item node $i$ as $\mathbf{Z}^u\in R^{|\mathcal{S}(u)|\times d}$ and $\mathbf{Z}^i\in R^{|\mathcal{S}(i)|\times d}$, where $\mathcal{S}(u)$ and $\mathcal{S}(i)$ are the intent node neighbor sets.

We first calculate two matrices: $\mathbf{P}^{u,i}\in R^{|\mathcal{S}(u)||\mathcal{S}(i)|\times 1}$, containing cosine similarities between each pair of rows in $\mathbf{Z}^u$ and $\mathbf{Z}^i$, and $\mathbf{D}^{u,i} \in R^{|\mathcal{S}(u)||\mathcal{S}(i)|\times d}$, containing $\mathbf{Z}^i_{[q,:]}-\mathbf{Z}^u_{[p,:]}$ for each row index $p=0,1,2,\cdots |\mathcal{S}(i)|-1$ and $q=0,1,2,\cdots |\mathcal{S}(u)|-1$.

Then, to enforce the relationship of more similar intent pairs inducing a smaller translation vector, the resulting vector $\mathbf{r}^{u,i}$ is computed via:

\begin{equation}
\label{eq:apply-IKGR}
    \mathbf{r}^{u,i}=\text{softmax}(\mathbf{P}^{u,i})^T\mathbf{D}^{u,i}
\end{equation}

\subsection{Intent Aware Scoring}
To further leverage signal from the intents extracted during KG construction, we incorporate them into the user-item interaction scoring function, in conjunction with the more traditional embedding similarity score. At inference time, given a tuple of user and item $(u, i)$, we derive \textit{embedding matching} and \textit{intent matching} scores between a user and an item, and the final score is a hybrid combination of them. 
The \textit{embedding matching} score between $u$ and $i$ is the cosine similarity between their embeddings:

\begin{equation}
\label{eq:eq1}
z_{u,i} = \frac{\mathbf{e}_u\cdot \mathbf{e}_i}{\parallel \mathbf{e}_u\parallel\cdot  \parallel \mathbf{e}_i\parallel }
\end{equation}

We denote a single intent extracted from item $i$ as $n^{(i)}$, and the collection of all intents for $i$ as $\Omega_{i}$. Similarly, a single intent extracted from $u$ and their collections is denoted as $n^{(u)}$ and $\Omega_{u}$.
Additionally, $\mathbf{e}_n$ is used to represent the embedding of some entity $n$.

The \textit{intent matching} score between $u$ and $i$ is determined by:

\begin{equation}
\label{eq:eq2}
y_{u,i} = \mathop{max}\limits_{n^{(u)}\in\Omega_u, n^{(i)}\in\Omega_i } \frac{\mathbf{e}_{n^{(u)}}\cdot \mathbf{e}_{n^{(i)}}}{||\mathbf{e}_{n^{(u)}}\parallel \cdot  \parallel \mathbf{e}_{n^{(i)}}\parallel }\cdot 0.5^{\mathbb{I}( \Omega_{u} \cap \Omega_{i} = \varnothing )}
\end{equation}

Equation \ref{eq:eq2} has two components, the similarity component and the non-overlap punishment component. In the event that $u$ and $i$ share a single intent (no need to have completely equal intent set), the non-overlap punishment component will be disabled (=1), otherwise if no intents are shared, it will punish by 0.5. The choice of 0.5 is decided based on empirical comparison along the development of the approach, which shows priority over too severe punishment values (e.g. 0.1).
Finally, the score for the triplet $(u, i)$ is the hybrid mixture of the embedding score and intent matching score:

\begin{equation}
\label{eq:final-score}
score(u,i) = y_{u,i} + \lambda z_{u,i}
\end{equation}
where $\lambda$ is an empirical mixer coefficient.
In the experiments, the coefficient $\lambda$ is set to 0.1 based on grid search result on the validation set.

In the ablation study section Section \ref{sec:exp-ablation}, 
\textit{Opt2. Vanilla GNN} means translation layers, simply use a two-layer GNN trained with standard ranking loss.
\textit{Opt3. Vanilla Trans.} means simply use a translation layer without GNN and the scoring function of Equation \ref{eq:final-score}. 
\textit{Opt4. Vanilla Scoring} means keep both GNN and translation layer, but removing intent-aware scoring function of Equation \ref{eq:final-score}.

\section{Related Works Continued}
\label{appendix:related works}

\textbf{Recommender Approaches Addressing Data Sparsity}.
Recent works tackle the data sparsity issue through enhanced model architectures, graph-based techniques, and knowledge-aware methods. The Item History Model (IHM) improves cold-start item recommendations by directly injecting user-interaction data into the item tower and employing an inductive structure for real-time inference \cite{zuo2024inductive}. Cold Brew addresses sparsity in graph neural networks by distilling node representations, mitigating the impact of missing or noisy neighbors \cite{zheng2021cold}. RippleNet further alleviates sparsity by propagating user preferences through knowledge graph relations, enriching item representations beyond collaborative signals \cite{wang2018ripplenet}. These approaches demonstrate that leveraging historical interactions, distilling structural knowledge, and integrating external information effectively mitigate data sparsity in recommendation systems.

\textbf{Directly Leveraging LLM as the Recommender}.
A-LLMRec enhances LLM-based recommendation by leveraging embeddings from state-of-the-art collaborative filtering models, excelling in both cold and warm scenarios while maintaining efficiency and model-agnostic integration \cite{kim2024large}. CoRAL introduces collaborative retrieval-augmented prompting, addressing LLMs' reliance on semantic information by incorporating user-item interactions through reinforcement learning-based retrieval policies, significantly improving long-tail recommendation \cite{wu2024coral}. TALLRec proposes a tuning framework to align LLMs with recommendation-specific tasks, demonstrating strong generalization and efficiency even with limited training data \cite{bao2023tallrec}. These approaches collectively highlight the potential of LLMs as standalone recommenders by addressing cold-start challenges, enhancing collaborative reasoning, and improving alignment with recommendation-specific objectives.

\textbf{Personalized Re-ranking Recommender System}.
Re-ranking in recommender systems aims to refine an initially ranked list to better capture user preferences and item relationships. Traditional ranking methods optimize global performance but often overlook the mutual influence between items and user-specific intent \cite{pei2019personalized}. Recent approaches address these limitations by integrating semantic tag information \cite{liang2010connecting}, multi-stage ranking optimization \cite{evnine2024achieving}, and item relationships \cite{liu2020personalized}. Transformer-based models effectively model global item interactions, while graph-based methods leverage item relationships for improved ranking \cite{pei2019personalized, liu2020personalized}. Furthermore, self-supervised learning enhances sequential recommendation by mitigating data sparsity issues \cite{zhou2020s3}. These methods collectively demonstrate that incorporating user personalization, item dependencies, and efficient ranking strategies significantly enhances re-ranking effectiveness.

\end{document}